\documentstyle[11pt,a4wide,epsf]{article}

\begin{document}

\begin{center}
{\Large \bf A Model for  Dwarf Spheroidal Satellite Galaxies Without Dark
Matter}\\[0.4cm]
{\large Ralf~S.~Klessen} \\[0.15cm]
{\small Max-Planck-Institut
f\"{u}r Astronomie, K\"onigstuhl 17, D-69117 Heidelberg, Germany}\\[0.20cm]
{\small e-mail: klessen@mpia-hd.mpg.de}
\end{center}

\vspace{0.5cm}
\begin{abstract}
Self-consistent simulations of the dynamical evolution of low-mass
satellite galaxies without dark matter on different orbits interacting
with an extended Galactic dark halo are described.  The calculations
proceed for many orbital periods until well after the satellite
dissolves.  In all cases the dynamical evolution converges to a remnant
that contains roughly 1~per cent of the initial satellite mass. The
stable remnant results from severe tidal shaping of the initial
satellite. To an observer from Earth these remnants look strikingly
similar to the Galactic dwarf spheroidal satellite galaxies.  Their
apparent mass-to-light ratios are very large despite the fact that
they contain no dark matter.

These computations show that a remnant without dark matter displays
larger line-of-sight velocity dispersions $\sigma$ for more eccentric
orbits, which is a result of projection onto the observational
plane. Assuming they are not dark matter dominated, it follows that
the Galactic dSph satellites with $\sigma >6$~km/s should have orbital
eccentricities of $e>0.5$. Some remnants have sub-structure along the
line-of-sight that may be apparent in the morphology of the horizontal
branch.
\end{abstract}

\section{Introduction}
\label{sec:intro}
At least about ten dwarf spheroidal (dSph) galaxies are known to orbit
the Milky Way at distances ranging from a few tens to a few hundred
kpc. On the sky they are barely discernible stellar density
enhancements. Some have internal substructure and appear flattened.
Their velocity dispersions and stellar masses are similar to those
seen in globular clusters. However, they are about two orders of
magnitude more extended.  For spherical systems in virial equilibrium
with an isotropic velocity dispersion, the overall mass of the system
can be determined from the observed velocity dispersion.  Comparing
this `gravitational' mass to the luminosity of the system determines
the mass-to-light ratio $M/L$, which for the solar neighborhood is $3
< M/L < 5$ (e.g. Tsujimoto et al. 1997).  For the dSph satellites,
$M/L$ values as large as a few hundred are inferred, leading to the
conclusion that these systems may be completely dark matter dominated
(see Da Costa 1998, Ferguson \& Binggeli 1994, Grebel 1997, Irwin \&
Hatzidimitriou 1995, Mateo 1998a,b, Pryor 1994).

An alternative might be that the assumption of virial equilibrium is
violated for the Galactic satellite galaxies: they could be
significantly perturbed by Galactic tides. Indeed, departures of the
dSph stellar density profiles from the best-fitting King models are
commonly interpreted as existence of `extra-tidal' stars implying that
most dSph galaxies could be significantly losing mass (Irwin \&
Hatzidimitriou 1995, Kuhn et al.~1996, Smith et al.~1997, Burkert
1997). However, this is still under debate (Olszewski 1998, Pryor
private communication).

The `tidal scenario' has been studied in detail by a variety of
authors: Oh et al. (1995) modeled the evolution of dSph galaxies on
different orbits in a set of rigid spherical Galactic potentials,
using $10^3$ particles for the satellites.  Piatek \& Pryor (1995)
concentrate on one perigalactic passage of a dSph galaxy in different
 Galactic potentials. Their satellite consists of
$10^4$ particles.  They find that a single perigalactic passage
cannot perturb a satellite significantly enough for an observer to
measure a high $M/L$ ratio, reaching similar conclusions as Oh et
al. (1995). Johnston et al. (1995) and Johnston (1997) are studying
the overall distribution of tidal debris from disrupted dwarf galaxies
in a Galactic halo.

\section{A Scenario for dSph Galaxies as Remnants of Dissolved
Satellites } 
\begin{figure}[ht]
\unitlength1cm
\begin{picture}(13,3)
\put(-1.10,-2.90){\epsfxsize=7.7cm \epsfbox{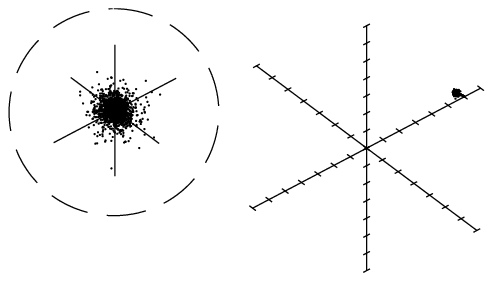}}
\put( 4.35,-2.90){\epsfxsize=7.7cm \epsfbox{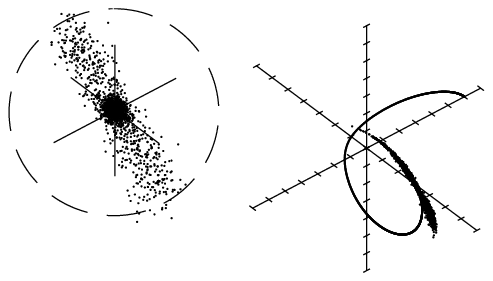}}
\put( 9.80,-2.90){\epsfxsize=7.7cm \epsfbox{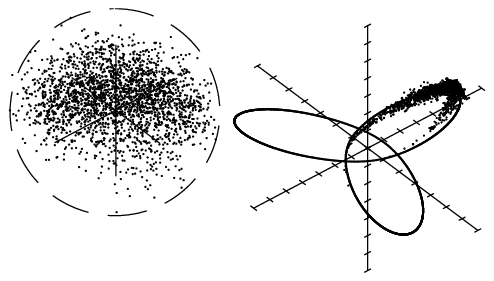}}
\end{picture}
\caption{\label{fig:1} {\small Snapshot of the evolution of a
satellite with orbital eccentricity of $e=0.71$.  The right side of
each panel plots the distribution of satellite particles at the given
time in the Galactic coordinate system. Each of the axes is 140$\:$kpc
long. The solid line depicts the trajectory of the density maximum of
the satellite until the time of the snapshot. Encircled are
enlargements of the central part of the satellite (the total length of
each axis is 5$\:$kpc).}}
\end{figure}

High-resolution simulations of the long-term evolution of
a low-mass satellite galaxy on different orbits interacting with an
extended Galactic dark halo are presented by Kroupa (1997; hereafter
K97) and by Klessen \& Kroupa (1998; hereafter KK98). The satellites
are modeled by $1.3\times 10^5$ up to $2.0\times 10^6$ particles and
the orbital eccentricities are in the range $0.41\le e\le 0.96$. The
simulations were performed using a particle-mesh code with nested
sub-grids and a direct-summation $N$-body code running with the
special purpose hardware device {\sc Grape} (Sugimoto et al. 1990,
Ebisuzaki et al. 1993).  Initially, the satellite is spherical with an
isotropic velocity distribution. It has a mass of $10^7\,{\rm
M}_\odot$ and a {\em true} mass-to-light ratio of $(M/L)_{\rm true} =
3.0$.  The calculations proceed for many orbital periods until well
after the satellite became unbound.  The aim is to study the system
well after the satellite has mostly dissolved and to `observe' its
properties as it would be seen from Earth. The satellite is projected
onto the sky and its brightness profile, line-of-sight velocity
dispersion and {\em apparent} $M/L$ ratio are determined.  These
quantities can be directly compared to the observed values for
Galactic dSph galaxies. Figure~\ref{fig:1} shows three snapshots of
the evolution of a typical satellite. The left panel depicts the
satellite immediately after the simulation started, the middle one
after the first apogalactic passage and the right one shortly after
the third apogalacticon. It has dissolved. However, there still exists
a measurable density enhancement, which might be identified as a dSph
galaxy.

\begin{figure}[ht]
\unitlength1cm
\begin{picture}(13,7.70)
\put( 1.00,-0.30){\epsfxsize=6.5cm \epsfysize=7.7cm \epsfbox{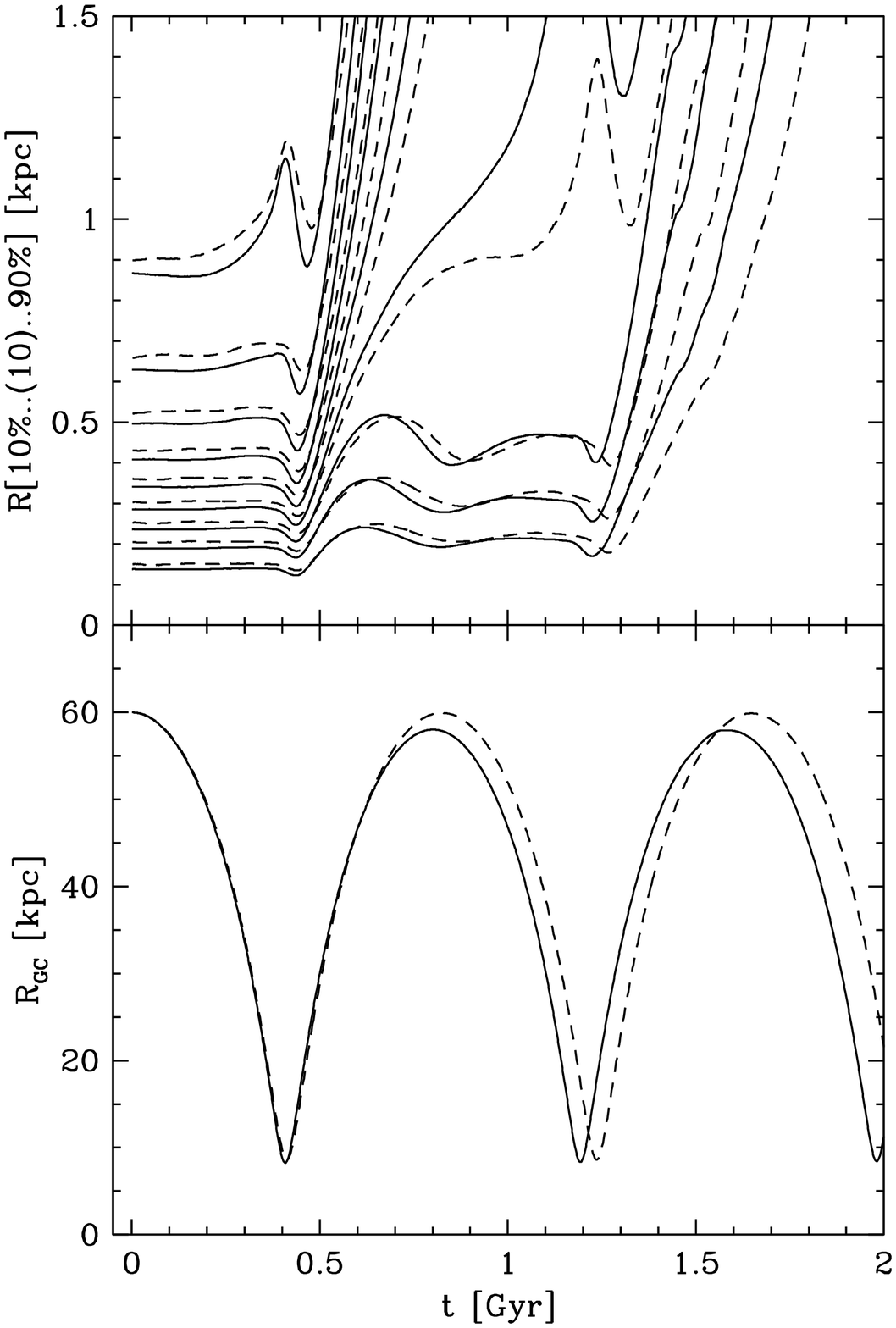}}
\put( 8.30,-0.30){\epsfxsize=5.7cm \epsfysize=7.7cm \epsfbox{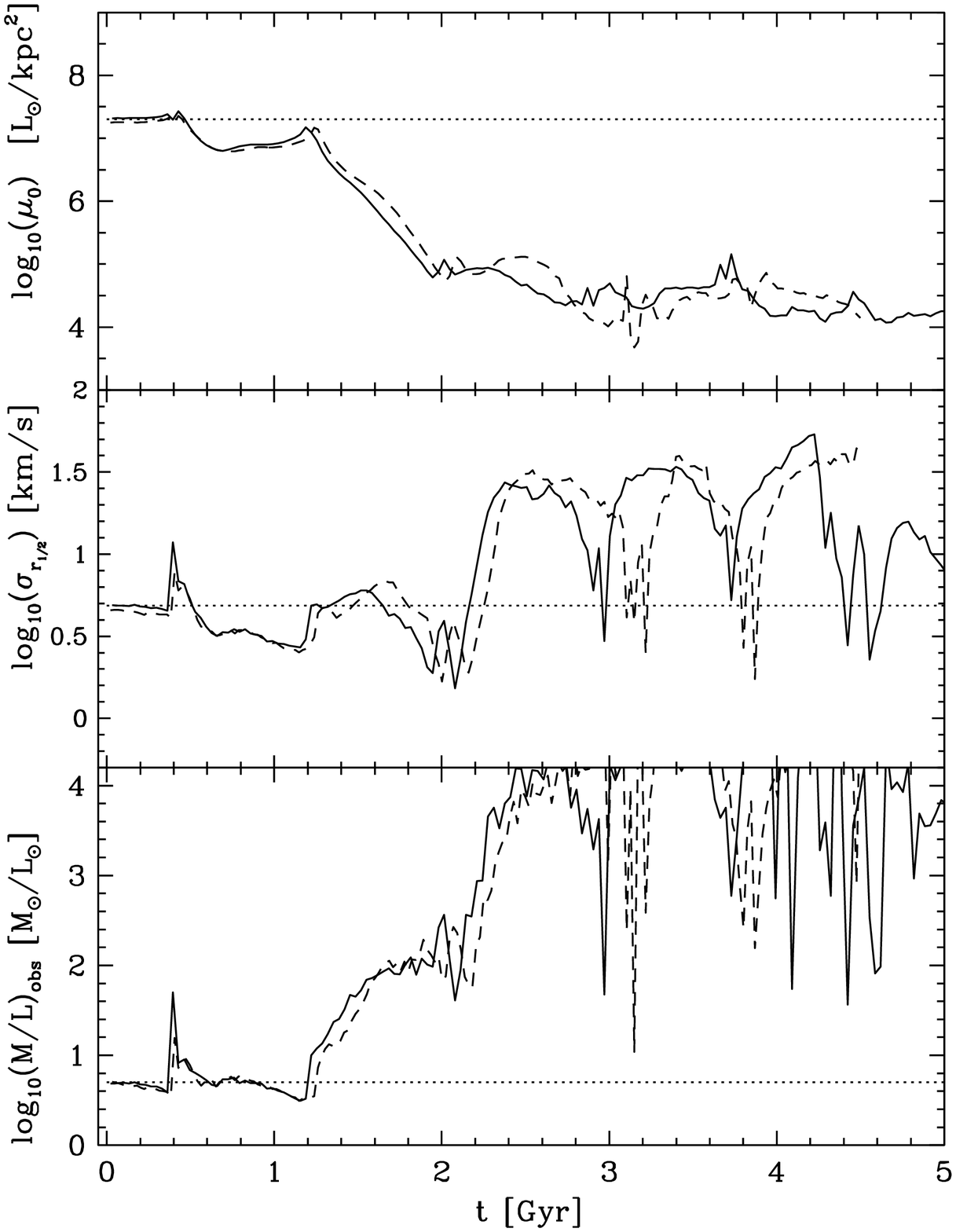}}
\end{picture}
\caption{\label{fig:2} {\small The left side shows the evolution of the radii
containing 10, 20, \dots, 90~per cent of the total mass of the
satellite (top) and the galactocentric distance (bottom) as a function
of time.  The right side plots the evolution of the central
surface brightness $\mu_0$ (top), of the line-of-sight velocity
dispersion $\sigma_{1/2}$ within the half-light radius (middle) and of
the apparent mass-to-light ratio (bottom). In all
panels, the solid line describes the simulation with the nested grid
code and the dashed line the one with direct summation on {\sc Grape}.
}}
\end{figure}

The main finding in this studies is that  a remnant containing about 1~per cent of the
initial satellite mass remains as a long-lived and distinguishable
entity after the major disruption event. To an terrestrial observer,
this remnant looks strikingly similar to a dSph galaxy.  The remnant
consists of particles that have phase-space characteristics that
reduce spreading along the orbit.    Projection
effects are also important: an observer who's line-of-sight subtends
a small angle with the orbital path of the satellite sees an apparently
brighter remnant with internal sub-clumps and an inflated velocity
dispersion.  The observer derives values for $(M/L)_{\rm obs}$ that are much
larger than the true mass-to-light ratio $(M/L)_{\rm true}$ of the
particles, because the object is far from virial equilibrium and has a
velocity dispersion tensor that is significantly anisotropic.
Figure~\ref{fig:2} shows in the left panel the evolution of the
Lagrange radii of the above satellite and the galactocentric distance
of its center-of-mass. The left panel plots the measured central
surface brightness $\mu$, the line-of-sight velocity
dispersion $\sigma$  and the derived mass-to-light ratio
$M/L$  as function of time. 

\section{Possible Discriminants}
The simulations in K97 and KK98 support the hypothesis that dark
matter may not be necessary to account for the structural and
kinematical properties of at least some of the Galactic dSph
satellites. They furthermore indicate possible diagnostics to
discriminate between tidal and dark-matter-dominated models.

\subsection{The Preference of Eccentric Orbits}
\begin{figure}[h]
\unitlength1cm
\begin{picture}(13,6.5)
\put( 1.50,-0.40){\epsfxsize=5cm \epsfbox{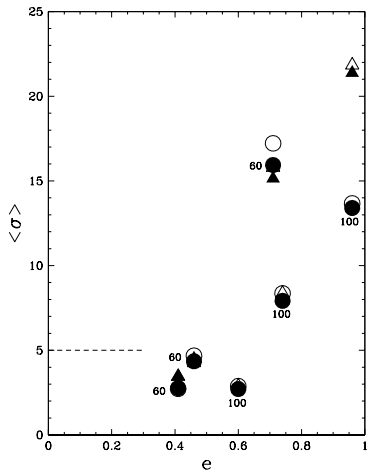}} \put( 8.00,
0.15){\parbox[b]{6.5cm}{\caption{\label{fig:3} {\small The time-averaged
velocity dispersion, computed over the first 2.5~Gyr after $(M/L)_{\rm
obs}\ge50$ is achieved, as a function of orbital eccentricity $e$ (from
KK98).  The horizontal dashed line indicates the initial central
line-of-sight velocity dispersion.  }}}}
\end{picture}
\end{figure}

The complete set of simulations discussed in K97 and KK98 shows that
there is a well-defined correlation between line-of-sight velocity
dispersion and orbital eccentricity. This can be quantified by
computing the time average of the observed central line-of-sight
velocity dispersion $\left<\sigma_0\right>$ over the time interval of
2.5 Gyr after the apparent mass-to-light ratio of the satellite has
exceeded the threshold $(M/L)_{\rm obs} = 50$. A plot of $\left<\sigma_0\right>$
as function of the eccentricity is given in
figure~\ref{fig:3}. Satellites initially on eccentric orbits lead to
apparently brighter remnants with inflated line-of-sight velocity
dispersions, owing to the observer's line-of-sight being approximately
aligned with the orbital path. In this case, particles ahead of and
following the remnant add to what the observer may make out to be a
dSph galaxy.  An observer looking along a very eccentric orbit finds a
remnant with a large $\sigma$.

The velocity dispersions measured in Galactic dSph satellites range
from about 6~km/s to~11~km/s (Irwin \& Hatzidimitriou 1995, Mateo
1998a).  The projection effects imply that, on
average, remnants without dark matter and with larger velocity
dispersions ought to be on more eccentric orbits and suggests that the
Galactic dSph satellites have orbital eccentricities
$e>0.5$. Conversely, a dSph satellite with $e<0.3$ and
$\sigma_{0}>6$~km/s should be dark matter dominated, unless
such it has an extreme internal velocity anisotropy with a
large velocity dispersion perpendicular to the direction of the
orbital path (Kuhn 1993).

\subsection{The Width of the Horizontal Branch}
\begin{figure}[ht]
\unitlength1cm
\begin{picture}(13,6.25)
\put( 2.00,-0.40){\epsfxsize=5cm \epsfbox{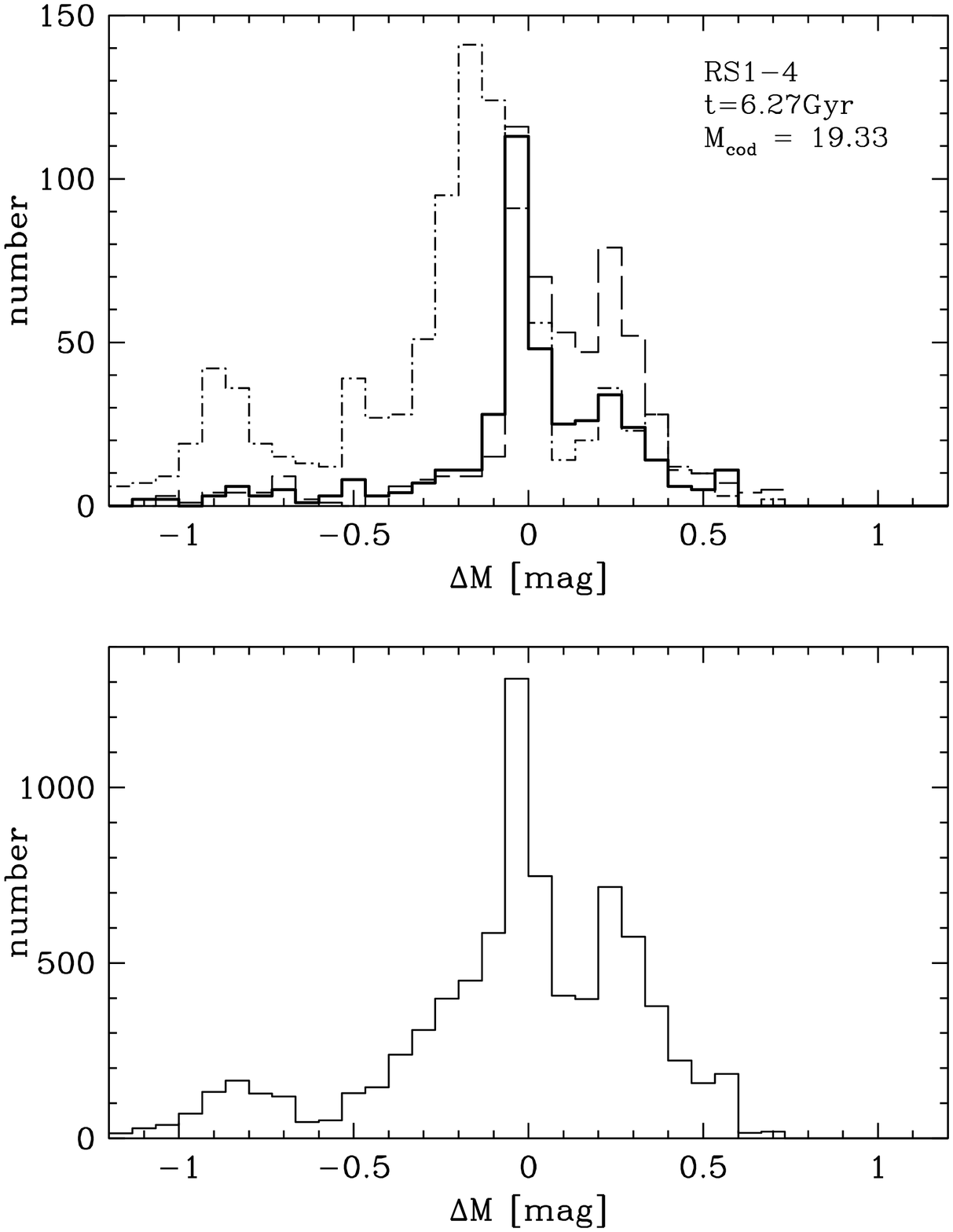}}
\put( 8.20,-0.40){\epsfxsize=5cm \epsfbox{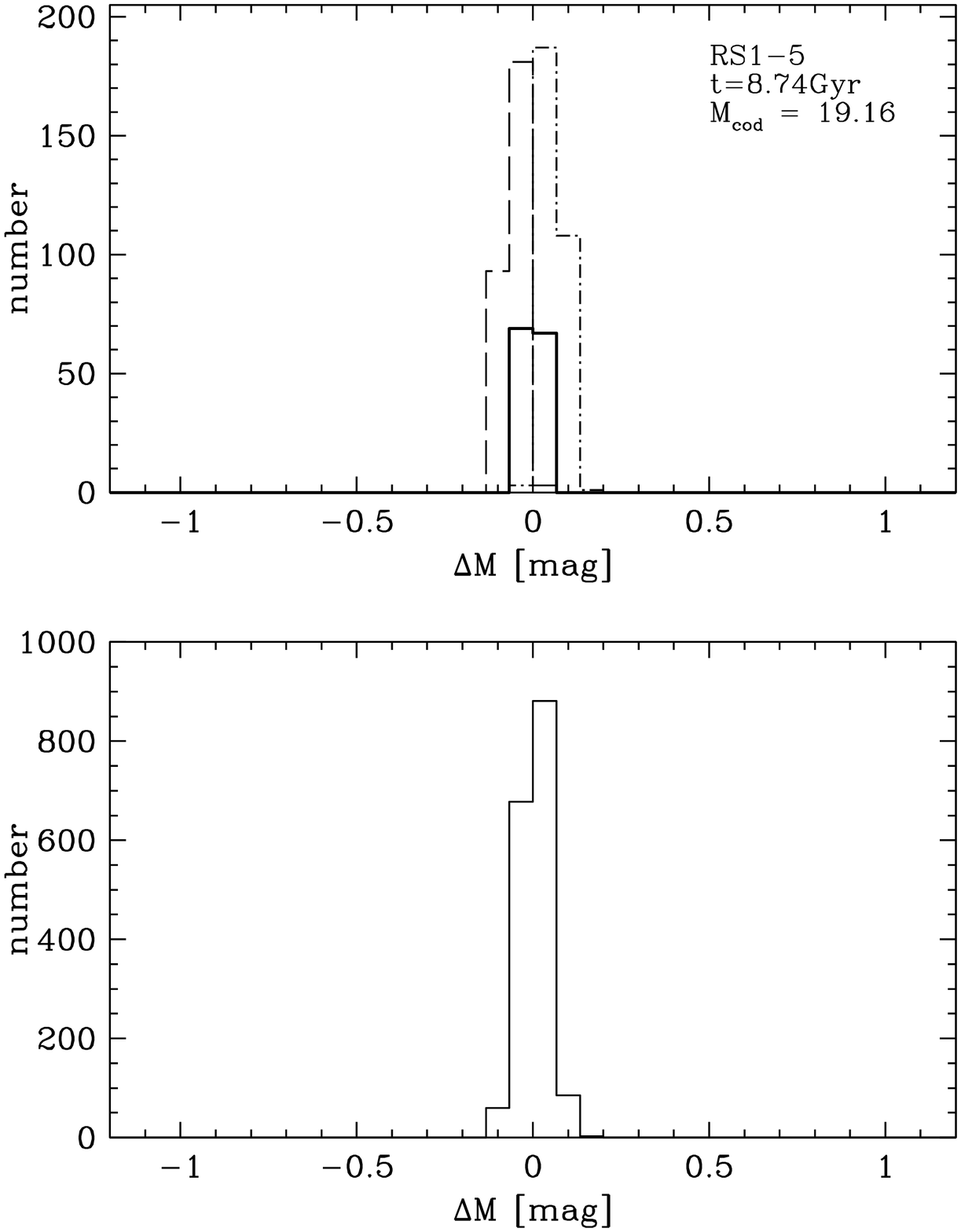}}
\end{picture}
\caption{\label{fig:4} {\small Upper panels: distribution of distance moduli
relative to the distance modulus of the remnant's density maximum at
three different positions across the face of the remnant. Lower panel:
distribution for all particles appearing projected within a radial
distance of 1.2~kpc from the position on the sky of the remnant's
density maximum. }}
\end{figure}

 A spread of distances leads to a broadening of the giant and
horizontal branches in the HR diagram.  Sub-clumping along the line-of
sight will lead to distinct populations that are separated vertically
in the HR diagram. These are important possible observational
discriminants, and the horizontal branch is especially well suited for
this type of investigation because it is horizontal and blue enough to
be less affected by contamination by foreground Galactic field stars.
The upper panels in figure~\ref{fig:4} show the apparent magnitude
distribution of the particles at three different positions across the
face of a remnant. The lower panels sample all particles within
1.2~kpc of the density maximum. In the left plot, the angle between
the line-of-sight and the orbital trajectory is very small.  The
radial extent of the remnant is very large and there is considerable
scatter in the diagram. On the right side, the remnant is seen almost
perpendicular to its orbital motion and its radial depth is very
small. Its HR diagram would appear very narrow.

The tidal model therefore predicts considerable scatter in the CMDs of
at least some dSph galaxies.  For preferentially eccentric orbits,
projection effects enhance the extent of the observed remnant along
the line-of-sight.  This translates into a distribution of stellar
magnitudes and subsequently a scatter in the CMD. This feature is
observed in some of the Galactic dSph galaxies (see Grebel 1997 and
Da~Costa 1997) and is usually interpreted as being a sign of complex
star formation histories or metallicity variations.  The above model
suggests an additional possibility. Probably all three contribute and
it seems very difficult to disentangle these effects.  \\[0.5cm]

\def\rfnce{\par\noindent\hangindent 20pt  {}}
\noindent  {\large \bf References}

{\small%
\rfnce{Burkert, A., 1997, ApJ, 474, L99}
\rfnce{Da Costa, G.S., 1997, in ``Stellar Astrophysics for the
      Local Group: A First Step to the Universe'', eds. A. Aparicio \&
      A. Herrero, Cambridge University Press, in press}
\rfnce{Ebisuzaki, T., Makino, J., Fukushige, T., Taiji, M., Sugimoto, D.,
      Ito, T., Okumura, S., 1993, PASJ, 45, 269}
\rfnce{Ferguson, H.C., Binggeli, B., 1994, A\&AR, 6, 67}
\rfnce{Grebel, E.K., 1997, Reviews in Modern Astronomy, 10, 29}
\rfnce{Grillmair, C.J., 1998, in the proceeding to the 1997 Santa
Cruz Halo Workshop, ed. D.~Zaritsky, in press (also astro-ph/9711223)}
\rfnce{Irwin, M., Hatzidimitriou, D., 1995, MNRAS, 277, 1354}
\rfnce{Johnston, K.V., 1997, ApJ submitted (also astro-ph/9710007)}
\rfnce{Johnston, K.V., Spergel, D.N., Hernquist, L., 1995, ApJ,
451, 598} 
\rfnce{Klessen, R.S., Kroupa, P., ApJ, 498, in press (also
astro-ph/9711350)}
\rfnce{Kroupa, P., 1997, New Astronomy, 2, 139 (K97)}
\rfnce{ Kuhn, J.R., 1993, ApJ, 409, L13}
\rfnce{Kuhn, J.R., Smith, H.A., Hawley, S.L., 1996, ApJ, 469, L93}
\rfnce{Mateo, M., 1998a, in: ``The Nature of Elliptical Galaxies'', 
eds. M. Arnaboldi, G.S. Da Costa, \& P. Saha, PASP, Vol. 116 
(also: astro-ph/9701158) }
\rfnce{Mateo, M., 1998b, ARAA, in press} 
\rfnce{Oh, K.S., Lin, D.N.C., Aarseth, S.J., 1995, ApJ, 442, 142}
\rfnce{Olszewski, E.W., 1998, in the proceeding to the 1997 Santa
Cruz Halo Workshop, ed. D.~Zaritsky, in press (also astro-ph/9712280)}
\rfnce{Piatek, S., Pryor, C., 1995, AJ, 109, 1071}
\rfnce{Pryor, C., 1994, in ``Dwarf Galaxies'', eds. G.~Meylan \& P.~Prugniel, 
ESO Conference and Workshop Proceedings No. 49, ESO, p.~323}
\rfnce{Smith, H.A., Kuhn, J.R., Hawley, S.L., 1998, in ``Proper
Motions and Galactic Astronomy'', ed. R.M.~Humphreys, PASP, in press}
\rfnce{Sugimoto, D., Chikada, Y., Makino, J., Ito, T., Ebisuzaki, T.,
Umemura, M., 1990, Nature, 345, 33}
\rfnce{Tsujimoto, T., Yoshii, Y., Nomoto, K., Matteucci, F.,
Thielemann, F.-K. \& Hashimoto, M., 1997, ApJ, 483, 228}
\rfnce{}
}

\end{document}